\documentclass[aps]{revtex4}
\newcommand{\bb}{\begin{eqnarray}}
\newcommand{\ee}{\end{eqnarray}}
\begin{document}
\title{{Hawking temperature from tunnelling formalism}}
\author{P. Mitra}\email{parthasarathi.mitra@saha.ac.in}
\affiliation{Saha Institute of Nuclear Physics\\
Block AF, Bidhannagar\\
Calcutta 700 064}
%\date{}
%\date{hep-th/0611265}
\begin{abstract}
It has recently been suggested that the attempt to understand
Hawking radiation as tunnelling across black hole horizons
produces a Hawking temperature double the standard value. 
It is explained here how one can obtain the standard value
in the same tunnelling approach.
\end{abstract}
\maketitle
\bigskip

A classical black hole has a  horizon  beyond which  nothing
can leak out. But there is a relation between the area
of the horizon and the mass (and other parameters like the charge)
indicating a close similarity \cite{BCH} with
the thermodynamical laws, thus allowing the definition of an entropy
and a temperature \cite{Bek}. 
This analogy was surmised to be of quantum origin and made  quantitative
after the theoretical discovery of radiation from black holes \cite{Hawk}. 
For a Schwarzschild black hole, the radiation, which is thermal, has a
temperature
\bb
T_H={\hbar\over 4\pi r_h}={\hbar\over 8\pi M},
\ee
where $r_h$ gives the location of the horizon in standard coordinates
and $M$ is the mass of the black hole.
This was derived by considering quantum massless particles in a
Schwarzschild background geometry. The derivation being quite
complicated, attempts have been made to understand the process of
radiation by other methods. In \cite{HH}, a path integral study
was made, and analytic continuation
in complex time used to relate amplitudes for particle emission and
absorption with the result
that the ratio of emission and absorption probabilities for energy $E$ is
given by
\bb
P_{emission}=\exp(-{E\over T_H})P_{absorption}\label{abs}.
\ee
This ``detailed balance'' relation provides further evidence
for the temperature $T_H$. Furthermore, the propagator in the
Schwarzschild background was shown \cite{HH} to have a periodicity
in the imaginary part of time with period $4\pi r_h= 8\pi M$,
again suggesting the same temperature. There is also an argument
involving a conical singularity on passing to imaginary time, which
can only be avoided if the standard Hawking temperature is chosen. 

Later, other attempts were made to understand the emission of particles
across the horizon as a quantum mechanical tunnelling process \cite{PW}.
The approach of using (\ref{abs}) was followed in \cite{pad}. Different
Hamilton - Jacobi treatments were used to reproduce the standard temperature 
$T_H$ \cite{zerbini}. Recently, however,
it has been pointed out \cite{anti} that this approach seems to produce a
temperature that is {\it double} the standard value $T_H$, which corresponds
to a halving of the period in imaginary time. This is
reminiscent of \cite{thooft}, where it was pointed out that the Hawking
temperature could be doubled with a different interpretation of the
gravitational field in quantum theory. However, such an interpretation
is not used in \cite{anti}. So it becomes necessary to try to resolve the
contradiction between this and the earlier analyses. 

A massless particle in the Schwarzschild background is described
by the Klein-Gordon equation
\bb
\hbar^2(-g)^{-1/2}\partial_\mu(g^{\mu\nu}(-g)^{1/2}\partial_\nu\phi)=0.
\ee
One expands
\bb
\phi=\exp(-{i\over\hbar}S+...)
\ee
to obtain to leading order in $\hbar$ the equation
\bb
g^{\mu\nu}\partial_\mu S\partial_\nu S=0.
\ee
If we use separation of variables to write, {\it provisionally},
\bb
S=Et+S_0(r),
\ee
the equation for $S_0$ becomes
\bb
-{E^2\over 1-\frac{r_h}{r}} + (1-\frac{r_h}{r})S_0'(r)^2=0
\ee
in the Schwarzschild metric. The formal solution of this equation is
\bb
S_0(r)=\pm E\int^r{dr\over  1-\frac{r_h}{r}}.
\ee
The sign ambiguity comes from the square root and corresponds to the
fact that there can be incoming/outgoing solutions. There is, furthermore,
a singularity at the horizon $r=r_h$, which has to be handled if one tries to find
a solution across it. 

One way to skirt the pole is to change $r-r_h$ to  $r-r_h-i\epsilon$.
This yields
\bb
S_0(r)=\pm E[r+r_h\cdot i\pi+r_h\int^r dr P(\frac{1}{r-r_h})],
\ee
where $P()$ denotes the principal value. For the outgoing solution,
\bb
S_{out}=Et-E[r+r_h\cdot i\pi+r_h\int^r dr P(\frac{1}{r-r_h})],
\ee
the imaginary part yields a decay factor $\exp (-\pi r_hE/\hbar)$ in the amplitude
and hence a factor $\exp (-2\pi r_hE/\hbar)$ in the probability.
This has been interpreted to signal a temperature \cite{anti}
\bb
{\hbar\over 2\pi r_h}=2T_H,
\ee
twice as big as the standard Hawking temperature.

This observation given in \cite{anti} may suggest that one should 
dump the original calculation \cite{Hawk}. However
that calculation has not been directly challenged, nor can one forget the
other arguments in support of the standard value of $T_H$, for example
the one involving the periodicity in imaginary time or the conical
singularity in  passing to imaginary time. So we have to see if it
is possible to make sense of the imaginary part of the above $S_0$
without doubling the Hawking temperature.

A point made in \cite{zerbini} is that $r$ is not the proper radial distance,
and ought to be replaced by $\sigma\approx2\sqrt{r_h(r-r_h)}$ before introducing
an $i\epsilon$. However, the use of this variable involves a different
kind of path and the evaluation of the integral by \cite{zerbini} has been criticized
in \cite{anti}. 

It is more interesting to compare the above argument with \cite{pad}, where,
following \cite{HH}, the principle of detailed balance
(\ref{abs}) is used. Instead of just looking at
the outgoing solution, one then has to consider the incoming solution as well:
\bb
S_{in}=Et+E[r+r_h\cdot i\pi+r_h\int^r dr P(\frac{1}{r-r_h})].
\ee
The imaginary part here yields a factor  $\exp (\pi r_hE/\hbar)$ in the amplitude, 
leading to a factor  $\exp (2\pi r_hE/\hbar)$ in the probability. The {\it ratio
of the outgoing and incoming probabilities} is  $\exp (-4\pi r_hE/\hbar)$,
which is as in (\ref{abs}). This is how one can think of 
obtaining the standard temperature instead of getting twice the
value. But curiously the above incoming factor is an
{\it amplification}, not a decay, so that 
the absorption probability tends to be 
greater than unity and goes to infinity in the classical limit.

Let us instead rewrite the outgoing and incoming solutions as
\bb
S_{out}&=&Et+C-E[r+r_h\cdot i\pi+r_h\int^r dr P(\frac{1}{r-r_h})],\nonumber\\
S_{in}&=&Et+C+E[r+r_h\cdot i\pi+r_h\int^r dr P(\frac{1}{r-r_h})],
\ee
where $C$ is the constant {\it arising from the integration of} 
${\partial S\over \partial t}=E$. 
The real part of the hitherto suppressed $C$ is indeed quite arbitrary, but
the imaginary part is not. It has to be determined so as to cancel the imaginary part
of $S_{in}$. This is essential to ensure that the incoming probability
is unity {\it in the classical limit} -- when there is no reflection and everything is 
absorbed -- instead of zero or infinity. Thus,
\bb
C&=&-i\pi r_hE + (Re~C),\nonumber\\
S_{out}&=&Et-E[r+r_h\cdot 2i\pi+r_h\int^r dr P(\frac{1}{r-r_h})]+ (Re~C),
\ee
implying a decay factor $\exp (-2\pi r_hE/\hbar)$ in the amplitude,
and a factor $\exp (-4\pi r_hE/\hbar)$ in the probability,
in conformity with the standard value of the Hawking temperature.

The above calculation has been done in Schwarzschild coordinates.
Alternative coordinates that have often been used for tunnelling studies
are the ones due to Painlev\'{e} \cite{PW,anti}. In this case the
calculation of $S$ \cite{anti} shows that  $S_{in}$ has no imaginary part to
begin with, so that there is no need to introduce a non-zero $Im~C$. One finds further
\bb
Im~S_{out}=- 2\pi  r_hE
\ee
directly, yielding the expected decay factor  $\exp (-4\pi r_hE/\hbar)$ 
in the probability. There is no lack
of consistency \cite{chow,anti} between the Schwarzschild
and the Painlev\'{e} formulations, and it is reassuring to note that
$Im~S_{out}-Im~S_{in}=- 2\pi  r_hE$ in both the Schwarzschild and the
Painlev\'{e} cases, irrespective of the value of the complex constant $C$.

Here we have restricted ourselves to the simplest black hole horizon. It is
easy to check that these ideas work also in the case of, say, the de Sitter 
horizon and the Rindler horizon (see the second paper in \cite{zerbini}). 

In short, there is no problem with the standard value of the Hawking temperature.
Hawking radiation at the standard temperature
can be understood through tunnelling, contrary to the view of \cite{anti}.
The crucial step is to note that the classical absorption probability is unity.

\acknowledgments
I thank Amit Ghosh for several discussions on this issue.

\end{document}